\documentclass[letterpaper]{article} 
\usepackage[draft]{aaai2027}  
\usepackage{times}  
\usepackage{helvet}  
\usepackage{courier}  
\usepackage[hyphens]{url}  
\usepackage{graphicx} 
\usepackage{amssymb}
\usepackage{natbib}  
\usepackage{caption} 
\usepackage[table]{xcolor}
\usepackage{booktabs}
\usepackage{array}
\usepackage{tabularx}
\usepackage{multirow}
\usepackage{makecell}
\usepackage{colortbl}
\usepackage{enumitem}

\usepackage{tikz}
\definecolor{botbg}{HTML}{14386F}
\definecolor{convbg}{HTML}{F0B646}
\newcommand{\bottag}[1]{\tikz[baseline=(t.base)]\node[fill=botbg,text=white,rounded corners=2.5pt,inner xsep=3.5pt,inner ysep=1.5pt,font=\scriptsize\sffamily](t){#1};}
\newcommand{\convnum}[1]{\tikz[baseline=(t.base)]\node[fill=convbg,text=botbg,rounded corners=2.5pt,inner xsep=3.5pt,inner ysep=1.5pt,font=\scriptsize\sffamily](t){#1};}
\newcommand{\botconvtag}[2]{\bottag{#1}\hspace{2pt}\convnum{#2}}

\usepackage{algorithm}
\usepackage{algorithmic}

\usepackage{newfloat}
\usepackage{listings}
\DeclareCaptionStyle{ruled}{labelfont=normalfont,labelsep=colon,strut=off} 
\floatstyle{ruled}
\newfloat{listing}{tb}{lst}{}
\floatname{listing}{Listing}
\title{Misalignment Between Educators' Testing and End Users' Behavior:\\
What 17,000+ Chatbot Messages Reveal About AI Literacy Training for Teachers}
\author {
    Ruiwei Xiao\textsuperscript{\rm 1},
    Jiayi Wang\textsuperscript{\rm 2},
    Miina Koyama\textsuperscript{\rm 1},
    Yue Wang\textsuperscript{\rm 1},\\
    Qing Xiao\textsuperscript{\rm 1},
    Xinying Hou\textsuperscript{\rm 3},
    Norman Bier\textsuperscript{\rm 1},
    John Stamper\textsuperscript{\rm 1}
}
\affiliations {
    \textsuperscript{\rm 1}Carnegie Mellon University\\
    \textsuperscript{\rm 2}University of Maryland\\
    \textsuperscript{\rm 3}University of Michigan\\
    ruiweix@andrew.cmu.edu
}

\begin{document}

\maketitle

\begin{abstract}
AI literacy training for educators is expanding rapidly, but its effectiveness rests largely on self-report, knowledge tests, or interviews, revealing little about whether learning carries into practice or affects students and colleagues who use what educators build. We implemented \textit{AI for Learning}, an asynchronous online course that trained 87 higher-education educators to recognize AI's capabilities, limitations, and ethics; apply prompt engineering; and evaluate AI outputs. Its central assignment asked educators to build, test, and voluntarily deploy chatbots for their practice on an educational chatbot authoring platform. Using LLM-assisted qualitative coding validated against human annotation, we analyzed the resulting 214 chatbots and their builder-bot, peer-bot, and end user-bot conversations, 17,000+ messages spanning 1.5 years. Educators 1) specified what bots should do more often than what they should not; 2) adopted templated prompt components but rarely untemplated strategies; and 3) tested routine use but seldom iterated, leaving 74\% of end users' adversarial use untested. Still, 36 bots (17\%) reached end users and 63 (29\%) were adopted by peers. We contribute two successes and two lessons spanning the lifecycle of educator AI literacy training.
\end{abstract}
\section{Introduction}

As generative AI enters everyday work and learning, AI literacy is becoming a basic skill for the general public \cite{long2020ailiteracy, ng2021conceptualizing}. It carries particular weight for educators, who now stand on both sides of the technology in the classroom. On the learner-facing side, chatbots configured with pedagogical guardrails have produced learning gains where unguarded ones produced losses \cite{kestin2025ai, bastani2025generative}, making teachers’ ability to specify and verify such guardrails consequential for what students get. On the teacher-facing side, AI literacy skills enable educators to use AI tools for lesson materials drafting, feedback generation, and polishing communication \cite{xiao2026teachers}. AI literacy professional development for educators has expanded accordingly, with initiatives such as an AI microcredential co-designed with K-12 teachers \cite{hutchins2025empowering}, a teacher-led AI literacy curriculum evaluated against researcher-led delivery \cite{zhang2024effectiveness}, and a TPACK-based program for computer science teachers \cite{sun2023tpackpd}.

Despite this general growth, evidence of what such professional training achieves remains difficult to gather. As programs are usually evaluated at course exit, the assessment almost always through what educators report about themselves. Self-report dominates, with attitude and confidence scales administered before and after a workshop \cite{younis2024effectiveness, lintner2024scales}. Knowledge tests add an objective check \cite{chiu2024developing, laupichler2022scoping}, but they test recall rather than performance on an authentic AI task, and self-reported and objective measures diverge even within the same cohort \cite{zhang2026misalignment}. Interviews supply depth \cite{hutchins2025empowering, le2026ailiteracytraining}, but they too end with the course. None of these instruments follows educators into practice or asks what happened to the students and colleagues who used what educators built. What is missing is evidence of what educators \textbf{\textit{actually do}} with AI after training, and of what that does for those around them.

We address both gaps through a single instrumented course. We implemented \textit{AI for Learning}, an asynchronous online course that trained 87 higher-education educators toward three learning objectives: recognizing AI's capabilities, limitations, and ethical considerations (LO1), applying prompt engineering skills (LO2), and evaluating AI-generated responses (LO3) (Section~\ref{sec:course-overview}). Its central assignment asked participating educators to build, test, and, if they chose, deploy AI chatbots for their own practice on Playlab, a no-code authoring platform. Playlab stores each chatbot's system prompt along with a full transcript of every conversation held with it, by its builder during testing and by anyone who chatted with it after deployment (e.g., fellow educators and end users). We coded the system prompts of the resulting 214 chatbots and the 1,740 conversations (over 17,000 messages) in which builders, peers, or end users interacted with them, using an LLM-assisted workflow validated against human annotation (Section~\ref{sec:analysis}).

We ask whether the learning and impact each objective targets are visible in this record: what educators wrote into their bots (LO1 and LO2), how they tested them (LO3), and what happened once users arrived. This work contributes two successes and two lessons spanning the lifecycle of educator AI literacy training, informing how future training can be designed, evaluated, and scaled.
\section{Related Work}

\subsection{AI Literacy Education for Teachers}

Frameworks for AI literacy specify what learners should understand about AI, spanning conceptual knowledge, critical evaluation, and ethical reasoning \cite{long2020ailiteracy, ng2021conceptualizing}. Teacher-facing formulations extend these competencies toward instructional judgment, asking educators not only to understand AI but to decide where it belongs in a lesson \cite{vandenberg2026co, zhou2025competency, celik2023aitpack}. A growing body of professional development work operationalizes such frameworks as workshops, courses, and modules for pre- and in-service teachers \cite{le2026ailiteracytraining, younis2024effectiveness, sun2023tpackpd}, and systematic reviews chart this rapidly expanding space \cite{tan2025artificial, guericson2025ailitreview}. Evaluation of the instruction itself, however, rests almost entirely on self-report: confidence scales, attitude surveys, and knowledge tests at module boundaries \cite{lintner2024scales, younis2024effectiveness, le2026ailiteracytraining, laupichler2022scoping}. Such measures capture what educators say they learned, not their performance on an authoring task; the artifacts educators produce during instruction remain an unexploited source of transfer evidence.

\subsection{End-User Authoring of AI Chatbots}

Natural-language prompting has made LLM behavior configurable without programming expertise, recasting prompt writing as end-user development \cite{zamfirescu2023johnny, wu2022aichains}. Studies of this practice report that non-experts underestimate specification effort, overgeneralize from single successful trials, and struggle to anticipate inputs unlike their own \cite{zamfirescu2023johnny, subramonyam2024gulf}. Systems work responds with tooling for systematic iteration and evaluation: side-by-side comparison and hypothesis testing \cite{arawjo2024chainforge}, criteria-based scoring \cite{kim2024evallm}, and simulated students for reviewing pedagogical agents \cite{jin2025teachtune}. Within education, researchers have examined how teachers approach chatbot design, documenting scoping decisions, prompt templates, and testing habits \cite{yoo2025howteachers, hedderich2024theatre}. This work illuminates teachers' design practices, but leaves open how those practices connect to subsequent testing and use by students and colleagues over time.

\subsection{Teacher-Authored Chatbots in Classrooms}

Once deployed, teacher-configured chatbots have been studied primarily through student interaction traces, characterizing conversation quality, task alignment, shortcut behavior, and how far student dialogue realizes the teacher's instructional intent \cite{liu2026tasd, han2023recipe, kim2024promptanalytics}. Orchestration research foregrounds the teacher's real-time role, examining monitoring, triage, and intervention as students work with AI-supported learning systems \cite{dillenbourg2013orchestration, holsteinaleven2022, jin2025whotohelp}. Systematic reviews of educational chatbots highlight the importance of interaction design and educational context for instructional value \cite{wollny2021arewethereyet, kuhail2023interacting, yan2024practical}. Across this literature, the teacher-authored prompt enters the analysis as a fixed input whose provenance is not examined: how it came to take its final form, and what the educator saw before releasing it, falls outside the observational window.
\section{Methods}
\label{sec:course}
\subsection{The \textit{AI for Learning} Online Course}
\label{sec:course-overview}
The \textit{AI for Learning} course is an educator-facing, non-technical, practice-oriented online course developed and launched by Open Learning Initiative\footnote{\url{https://oli.cmu.edu/}}. Offered here as a self-paced asynchronous pilot, it is designed for completion in nine weeks at roughly four hours per week across four modules. Each module addresses 1-2 learning objectives (LOs) through short videos and readings, practice questions (multiple-choice and open-ended), case studies, and a learning-by-doing project using educational AI tools. The chatbot authoring activity analyzed here concludes Module~2 and uses Playlab,\footnote{\url{https://playlab.ai}} a nonprofit no-code platform on which educators build and share custom AI chatbots by writing a natural-language system prompt. By that point, participants had completed instruction on 3 LOs:

\begin{enumerate}[label=\textbf{LO\arabic*}, leftmargin=*]
  \item \label{lo:recognize} \textbf{Recognize AI as a new partner.}
        Examine AI's capabilities, limitations, ethical
        considerations (e.g., privacy concerns, policies for effective integration).
  \item \label{lo:prompt} \textbf{Apply prompt engineering skills.}
        Develop and refine skills to align curriculum, learning objectives,
        assessments, and instructional content.
  \item \label{lo:evaluate} \textbf{Evaluate AI-generated responses.}
        Use specific criteria to assess the quality and effectiveness of AI
        outputs in educational contexts.
\end{enumerate}

\subsection{Participants}
Eighty-seven educators working in higher education enrolled in the course and authored at least one AI chatbot during the end of module activity, producing 214 bots in total ($\mathit{Median} = 1$ per educator, range 1-13). Of these bots, 36 (17\%) were deployed to end users, and 63 (29\%) were used by other course participants. Because the course targeted higher-education settings, the end users of these bots were college students, instructors, or staff, and were therefore adults, as were all enrolled educators.

\begin{table*}[t]
  \centering
  \scriptsize
  \setlength{\tabcolsep}{4pt}
  \renewcommand{\arraystretch}{1.2}
  \begin{tabular}{@{}>{\centering\arraybackslash}m{1.5cm}
    >{\raggedright\arraybackslash}m{3cm}
    >{\raggedright\arraybackslash}m{8cm}
    >{\centering\arraybackslash}m{2cm}
    >{\centering\arraybackslash}m{2cm}}
    \toprule
     & \textbf{Data coded} & \textbf{Codes} &
      \makecell{\textbf{Human--Human $\kappa$}\textsuperscript{a}} &
      \makecell{\textbf{Human--LLM $\kappa$}\textsuperscript{b}} \\
    \midrule
    \makecell[c]{\textbf{LO1}\\Recognize AI} &
      \cellcolor{gray!10}214 system prompts &
      \cellcolor{gray!10}
      {\hangindent=1.5em\hangafter=1 \textit{\textbf{Capability}}: content generation, tutoring and coaching, feedback and improvement, personalization, just-in-time assistance, analytics and prediction\par}
      {\hangindent=1.5em\hangafter=1 \textit{\textbf{Limitation}}: hallucination awareness, direct-answer awareness, accuracy and scope\par}
      {\hangindent=1.5em\hangafter=1 \textit{\textbf{Ethics}}: confidentiality, diversity and inclusion\par} &
      \cellcolor{gray!10}0.77 & \cellcolor{gray!10}0.80 \\
    \makecell[c]{\textbf{LO2}\\Prompt\\engineering} &
      214 system prompts &
      {\hangindent=1.5em\hangafter=1 \textit{\textbf{Component}}: role, context, style, workflow, guardrail\par}
      {\hangindent=1.5em\hangafter=1 \textit{\textbf{Strategy}}: zero-shot, few-shot, chain-of-thought\par} &
      0.74 & 0.85 \\
    \multirow[c]{2}{1.8cm}[-\baselineskip]{\centering\makecell[c]{\textbf{LO3}\\Evaluation}} &
      \cellcolor{gray!10}713 builder-side conversations &
      \cellcolor{gray!10}
      {\hangindent=1.5em\hangafter=1 \textit{\textbf{Testing mode}}: iterative refinement, limited evaluation, opportunistic exploration, no testing\par} &
      \cellcolor{gray!10}0.93 & \cellcolor{gray!10}0.89 \\
    & 17,407 builder- and end-user messages &
      {\hangindent=1.5em\hangafter=1 \textit{\textbf{User message}}: desired, adversarial, others\par}
      {\hangindent=1.5em\hangafter=1 \textit{\textbf{Bot message}}: success, fail, others\par} &
      0.89 & 0.71 \\
    \bottomrule
  \end{tabular}
  \caption{Coding scheme and reliability by learning objective. LO1 and LO2
  codes were assigned in a single LLM call per bot and verified together;
  LO3 was coded at two levels, one call per bot for testing mode and one
  call per conversation for message-level codes. Human--human $\kappa$ is
  inter-rater agreement between two authors on the seed sample; human--LLM
  $\kappa$ is agreement between the LLM and one author on the stratified
  verification sample. Message-level $\kappa$ is pooled across user and bot
  codes. \textsuperscript{a}Computed on a 10\% seed sample.
  \textsuperscript{b}Computed on a 10\% verification sample.}
  \label{tab:analysis}
\end{table*}

\subsection{Data Collection}
The bot-authoring activity was completed in Playlab. For each bot, we exported its full configuration from Playlab: title, description, system prompt, and platform settings (model, interaction style, enabled tools). We further exported all conversations on each bot with per-message timestamps and speaker roles, distinguishing \emph{builder-bot} conversations (the pre-deployment conversations between the authoring educator and the bot), \emph{peer-bot} conversations (the conversations fellow course participants have with others' bots at any stage of their lifecycle), and \emph{end user-bot} conversations (the conversations anonymous end users, the
bot's target audience, have with the bot through its shared link). In total, we collected 214 bots from 87 unique builders and 1,740 conversations comprising 17,407 messages: 713 builder-bot conversations, 125 peer-bot conversations, and 902 end user-bot conversations spanning 1.5 years from February 2025 to July 2026. All data were de-identified prior to analysis, and the study was approved by our institution's IRB. All educators enrolled in the course and the end users of the bots they authored were adults.

\begin{figure*}[!t]
\centering
\includegraphics[width=.9\textwidth]{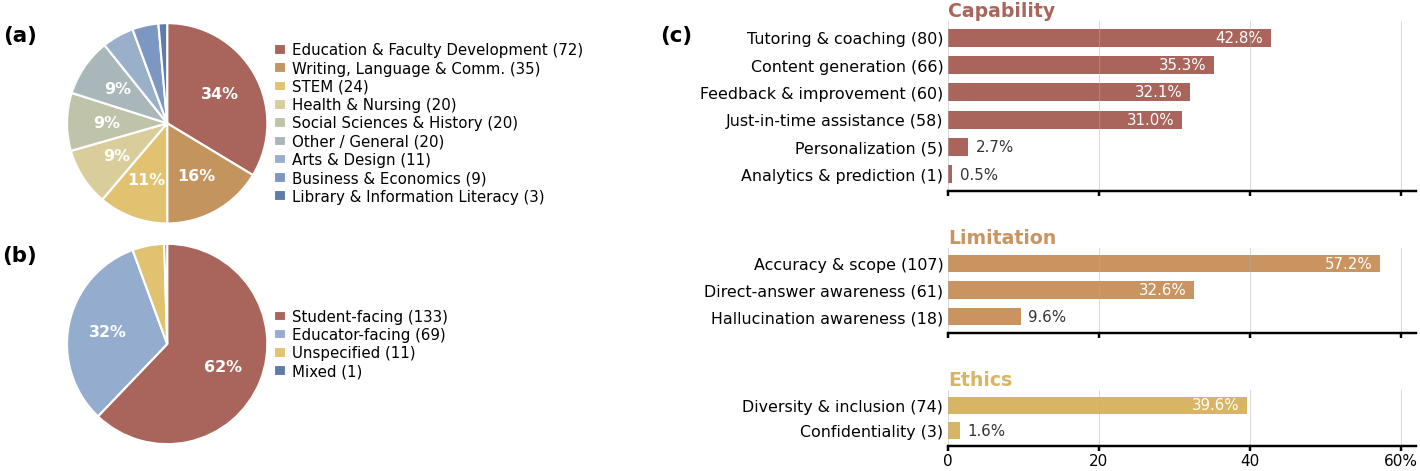}
\caption{LO1 overview: (a) subject domains, (b) intended audiences of all 214 bots, and (c) capability, limitation, and ethics code prevalence across 187 educator-authored prompts, excluding 27 unedited scaffold templates. Percentages in (a-b) use all 214 bots, and the bars in (c) use the 187 authored prompts. Parentheses give counts. A prompt may contain multiple codes.}
\label{fig:lo1-overview}
\end{figure*}

\subsection{Analysis}
\label{sec:analysis}
\subsubsection{LLM-Assisted Qualitative Coding.} Given the impracticality of manually coding 214 bots' metadata and 17,407 messages along 29 codes, we use LLM-assisted coding to label system prompts and chatbot conversations with authors or end users. This approach is increasingly adopted in mixed-methods research \cite{xiao2023deductive, chew2023llmassisted, ziems2024css, ma2026notwins} and reaches human-comparable reliability when paired with codebook-based prompting and human verification \cite{ziems2024css, chew2023llmassisted}. We follow the codebook-then-LLM-annotation workflow of \citet{xiao2023deductive, ma2026notwins}, verifying a stratified sample by hand and meeting the corresponding reliability bars on inter-rater agreement, annotation stability, and LLM-human agreement. Table~\ref{tab:analysis} summarizes data, codes, human--human and human--LLM coder reliability by LO.

We developed the codebook using a hybrid deductive-inductive approach \cite{fereday2006hybrid}. Initial codes drew deductively on course materials (capabilities, limitations, and ethics for \ref{lo:recognize}, and prompting components and strategies for \ref{lo:prompt}) and prior work on prompt authoring and evaluation for \ref{lo:evaluate} \cite{zamfirescu2023johnny, arawjo2024chainforge}. Two authors independently coded a random 10\% sample (22 bots, 171 conversations) using the draft, then resolved disagreements and revised the codebook by adding, merging or dropping codes. Inter-rater reliability was Cohen's $\kappa = 0.77$ for \ref{lo:recognize}, $0.74$ for \ref{lo:prompt}, and $0.93$ and $0.89$ for \ref{lo:evaluate} (testing mode and message codes). We froze the codebook before full-scale annotation.

LLMs then annotated all 214 bots, 1,740 conversations and 17,407 messages. Each bot's system prompt was labeled in a single call covering all \ref{lo:recognize} and \ref{lo:prompt} aspects, since several boundary rules adjudicate between them. \ref{lo:evaluate} was coded at two levels: one call per bot assigned a testing mode from all builder conversations with that bot, and one call per conversation labeled every user and bot message, chunking long conversations with preceding turns for context. The annotation prompt contained the full codebook with per-code definitions, examples, and exclusion rules, and required structured JSON output with a per-code rationale quoting the triggering phrase \cite{ma2026notwins}. We annotated with \colorbox{gray!10}{\texttt{Anthropic Claude Sonnet 4.5}} at temperature 0. Before the full run, we annotated the 10\% seed sample three times. Codes whose labels flipped across identical runs \cite{chew2023llmassisted} were flagged for oversampling in human verification.

One coder verified a stratified 10\% LLM-annotation sample, oversampling rare or unstable codes. LLM--human agreement (Cohen's $\kappa$) was $0.80$ for \ref{lo:recognize}, $0.85$ for \ref{lo:prompt}, $0.89$ for \ref{lo:evaluate} testing modes and $0.71$ for \ref{lo:evaluate} message codes, indicating good agreement \cite{landis1977measurement}.
\definecolor{resultnavy}{RGB}{128, 38, 48}
\definecolor{lo3navy}{RGB}{55, 90, 135}

\section{Results}

\subsection{LO1: Educators specified capabilities more consistently than limitations and ethics}
\label{sec:lo1}

LO1 asks educators to recognize AI as a new partner by examining its capabilities, limitations, and ethical considerations. We examine this learning outcome through what educators wrote into their bots' system prompts: which capabilities they drew on, and whether they also specified what the bot should not do. The 214 bots covered a broad range of subject domains (Figure~\ref{fig:lo1-overview}a). Education and faculty development was the largest category (72 bots, 34\%), followed by writing, language, and communication (35, 16\%) and STEM (24, 11\%). Most bots were designed for students (133, 62\%), while 69 (32\%) were educator-facing. The remaining bots had unspecified (11, 5\%) or mixed (1, $<1\%$) audiences (Figure~\ref{fig:lo1-overview}b). We coded each bot against 6 dimensions under capability, 3 under limitation, and 2 under ethics as described in Section~\ref{sec:analysis}. Of the 214 coded bots, 27 retained the unedited scaffold template and were excluded from prevalence estimates, leaving 187 authored prompts. Figure~\ref{fig:lo1-overview}c shows code prevalence across these 187 prompts.

\subsubsection{All prompts specify what the bot should do.} Every filled prompt named at least one AI capability, and prompts averaged 1.44 of the six coded capabilities. Tutoring and coaching was the most common (42.8\%), followed by content generation (35.3\%), feedback and improvement (32.1\%), and just-in-time assistance (31.0\%). Personalization (2.7\%) and analytics or prediction (0.5\%) were almost absent.

\subsubsection{Far fewer prompts specify what it should not do.} Attention to
limitations was less consistent. 71.1\% of prompts addressed at least
one limitation, most often by constraining accuracy or scope (57.2\%).
Far fewer instructed the bot to withhold direct answers (32.6\%) or
acknowledged hallucination (9.6\%). Ethical considerations were rarer still. Only 40.1\% of prompts contained any ethics code, nearly all of them diversity and inclusion (39.6\%). Confidentiality appeared in 1.6\%, and 34 prompts (18.2\%) addressed neither limitations nor ethics.

\begingroup
\color{black}

\definecolor{hi}{HTML}{2E7D32}\definecolor{mid}{HTML}{C77700}\definecolor{lo}{HTML}{B3261E}
\newcommand{\pcthi}[1]{\textcolor{hi}{#1}}
\newcommand{\pctmid}[1]{\textcolor{mid}{#1}}
\newcommand{\pctlo}[1]{\textcolor{lo}{#1}}

\begin{table*}[t]
\color{black}
\centering
\footnotesize
\renewcommand{\tabularxcolumn}[1]{m{#1}}
\begin{tabularx}{\textwidth}{@{}ll>{\raggedright\arraybackslash}X>{\raggedright\arraybackslash}m{0.32\textwidth}r@{}}
\toprule
\makecell{\textbf{Primary}\\\textbf{Code}} &
\makecell{\textbf{Secondary}\\\textbf{Code}} &
\textbf{Definition} & \textbf{Example} &
\makecell{\textbf{Prompts/}\\\textbf{Builders}} \\
\midrule
\multirow{5}{*}{Component}
  & \cellcolor{gray!10}Context
  & \cellcolor{gray!10}Background about the users and setting: who they are,
    their level, and their course
  & \cellcolor{gray!10}``You are talking to first-year students in MKT101''
  & \cellcolor{gray!10}\pcthi{90\%}\,/\,\pcthi{94\%} \\
  & Role
  & An explicit identity assigned to the bot
  & ``You are an expert in middle-school algebra''
  & \pcthi{88\%}\,/\,\pcthi{92\%} \\
  & \cellcolor{gray!10}Workflow
  & \cellcolor{gray!10}A sequence of steps the bot should follow, with at
    least two substantive steps
  & \cellcolor{gray!10}``First ask what topic they are studying; after they
    respond, offer a practice question''
  & \cellcolor{gray!10}\pcthi{87\%}\,/\,\pcthi{91\%} \\
  & Style
  & Educator-written tone, format, or length instructions
  & ``Use a friendly, encouraging tone and short paragraphs''
  & \pctmid{66\%}\,/\,\pctmid{66\%} \\
  & \cellcolor{gray!10}Guardrail
  & \cellcolor{gray!10}Educator-written prohibitions or scope fences on
    behavior outside the task
  & \cellcolor{gray!10}``Never give the final answer directly''
  & \cellcolor{gray!10}\pctmid{50\%}\,/\,\pctmid{57\%} \\
\midrule
\multirow{2}{*}{Strategy}
  & Few-shot
  & One or more worked examples embedded for the bot to imitate
  & ``Example 1: Q: \ldots{} A: \ldots{} Now respond in this style''
  & \pctlo{31\%}\,/\,\pctlo{33\%} \\
  & \cellcolor{gray!10}CoT
  & \cellcolor{gray!10}An instruction to reason in visible steps before answering
  & \cellcolor{gray!10}``Think aloud step by step as you evaluate the rubric''
  & \cellcolor{gray!10}\pctlo{2\%}\,/\,\pctlo{6\%} \\
\bottomrule
\end{tabularx}
\caption{LO2 codes across 214 system prompts by 87 builders, with
the definition and an illustrative example for each. Prompts/Builders
gives the share of prompts containing the code and the share of
builders using it in at least one of their prompts. All codes received
the same course instruction, and components additionally have slots or
stock lines in the starter template prompt. Only educator-authored
content counts toward style and guardrail. Zero-shot and
meta-prompting are omitted. Green $\geq$70\%, yellow 50--69\%, and
red $<$50\%.}
\label{tab:lo2}
\end{table*}

\subsection{LO2: Educators filled in the prompting template's elements, but skipped the unscaffolded prompt engineering strategies.}
\label{sec:lo2}

LO2 asks educators to apply prompt engineering skills to align curriculum, learning objectives, assessments, and instructional content. We examine this learning outcome through whether the prompt components and strategies taught in the course appear in the prompts educators wrote afterward. The course taught prompting through video tutorials and worked examples covering both prompt \emph{components} (role, context, style, workflow, guardrails) and prompting \emph{strategies} (zero-shot, few-shot, chain-of-thought, meta-prompting). \emph{Components} received one further support the \emph{strategies} did not: a starter template, introduced in the course and built into Playlab, with a slot or a stock line for each (e.g., ``You are an expert in \_\_\_''). To assess uptake, we labeled the system prompts of the chatbots educators built afterward (Section~\ref{sec:analysis}). Table~\ref{tab:lo2} reports how often each code appears across the 214 educational chatbots.

\subsubsection{Most educators covered the main \emph{components} the template provided.} Role (88\%), context (90\%), and workflow (87\%) are near-universal. Style (66\%) and guardrails (50\%) appear less often. Rather than being omitted, these components are typically left exactly as the template supplied them. Among prompts without an educator-authored style instruction, 96\% still carry the stock line (``Provide clear and concise responses'') unchanged, with no additions reflecting their own instructional setting or learner population. Additionally, 13\% of prompts still contain unfilled blanks, and 1\% of these bots were released to end users anyway.

\subsubsection{Prompting \emph{strategies} were rarely adopted without further support.} Few-shot prompting, embedding one or more worked examples in the prompt for the bot to imitate, appears in 31\% of prompts (33\% of builders). Chain-of-thought prompting, instructing the bot to reason in visible steps before answering, appears in five prompts (2\%). Meta-prompting, using AI to generate or refine the prompt itself, is not directly observable from prompt text and is excluded (Section~\ref{sec:analysis}).

\begin{table*}[!t]
  \color{black}
  \centering
  \small
  \setlength{\tabcolsep}{3pt}
  \renewcommand{\arraystretch}{1.15}
  \begin{tabular}{@{}>{\raggedright\arraybackslash}p{2.2cm} >{\raggedright\arraybackslash}p{4.8cm} >{\raggedright\arraybackslash}p{7.1cm} r r@{}}
    \toprule
    \textbf{Testing mode} & \textbf{Definition} & \textbf{Example builder turn(s)} & \textbf{Bots (\%)} & \textbf{Builders (\%)} \\
    \midrule
    \rowcolor{gray!10}
    Iterative refinement &
      Builder repeatedly tests and refines the bot based on conversation outcomes. &
      \botconvtag{Course Policy Assistant}{Conversation 1} \newline ``are you sure that objective 5 and 6 are not as important?'' \newline \botconvtag{Course Policy Assistant}{Conversation 2} \newline ``but are there any learning objectives that are more important than others?'' &
      30 (14\%) & 20 (23\%) \\
    Limited evaluation &
      Builder runs a small number of tests without deep iteration. &
      \botconvtag{ESL Syllabus Helper}{Conversation 1} \newline ``when is the first day of class?'' (acting as a student) \newline \botconvtag{Campus Resource Navigator}{Conversation 3} \newline ``hola'' (checking the bot's ``respond in the language written'' rule) &
      61 (29\%) & 33 (38\%) \\
    \rowcolor{gray!10}
    Opportunistic exploration &
      Builder explores the bot casually or inconsistently. &
      \botconvtag{Physics Tutor Bot}{Conversation 6} \newline ``teach me something'' \newline \botconvtag{Art Detective Bot}{Conversation 1} \newline ``ok'' \ldots\ ``now what?'' &
      83 (39\%) & 20 (23\%) \\
    No testing &
      Little or no builder testing evidence for this bot. &
      No substantive builder turn recorded &
      40 (19\%) & 14 (16\%) \\
    \bottomrule
  \end{tabular}
  \caption{Testing modes per bot (n = 214), adapted from
  \citet{arawjo2024chainforge}. Builders (\%) is the share of the 87 builders
  whose most rigorous mode across their own bots is this one.
  Tags give a descriptive pseudonym for the bot, assigned by
  the authors in place of the educator's original title, and the
  conversation's position in the builder's chronological sequence.}
  \label{tab:lo3-modes}
\end{table*}

\begin{figure*}[!t]
  \centering
  \includegraphics[width=\textwidth]{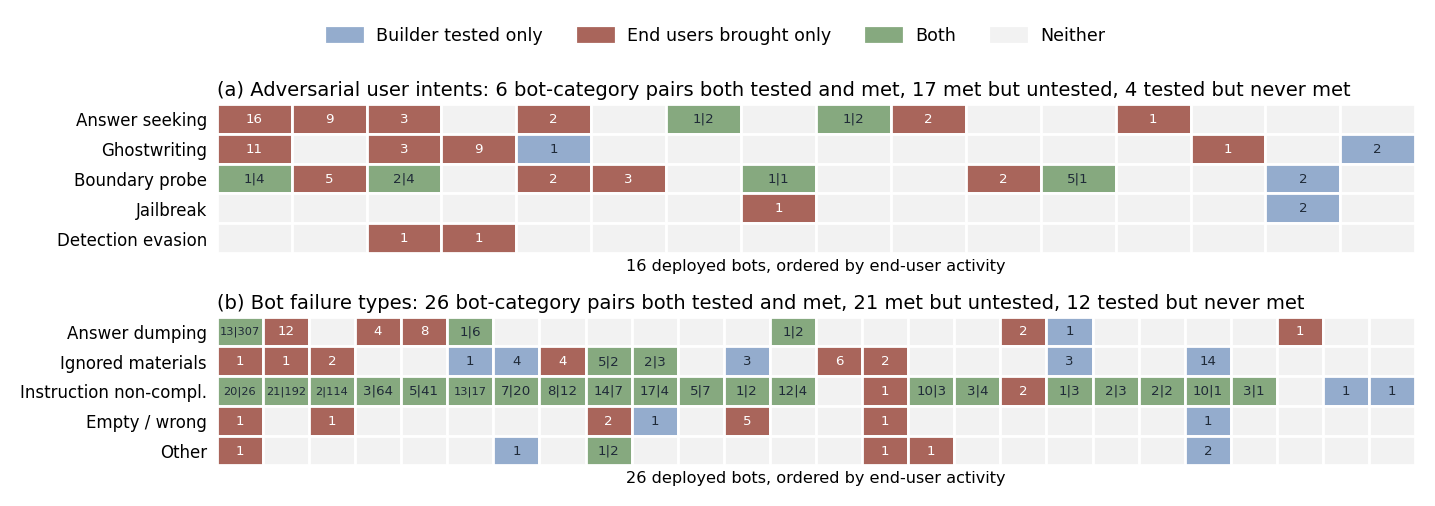}
  \caption{Builder--end-user overlap in (a) adversarial intents and (b) bot failures. Columns are deployed bots with at least one occurrence in each panel, ordered by end-user activity. Blue: builder only; red: end user only; green: both; gray: neither. Green cells show the \textit{number of builder-test messages / number of end-user messages}.}
  \label{fig:lo3-perbot-coverage}
\end{figure*}

\subsection{LO3: Routine use was tested often; iterative refinement and adversarial use were not}
\label{sec:lo3}

LO3 asks educators to evaluate AI-generated responses. We examine this learning outcome in two ways: how educators tested the bots they built, and whether that testing covered how end users later used them.

\subsubsection{Most educators tested their bots, but few tested systematically and iteratively.}

Of the 214 bots, 177 (83\%) had at least one builder-side conversation. These bots were built by 75 of the 87 educators (86\%). However, testing was limited and uneven. The median educator held three builder-side conversations ($\mathit{Median}=3$, $\mathit{IQR}=2$-$8$, $\max=99$), while the five most active educators accounted for half of all 713 builder-side conversations. Testing was also concentrated in time. For 61\% of tested bots, all builder-side conversations happened within a single day and before deployment.

We coded bot testing using the modes in Table~\ref{tab:lo3-modes}, adapted from \citet{arawjo2024chainforge}. At the bot level, 14\% showed iterative refinement, 29\% received limited evaluation, 39\% were explored opportunistically, and 19\% showed no substantive testing. Only 23\% of educators reached iterative refinement on at least one bot. Another 38\% conducted limited evaluation, usually checking the bot's main function or a behavior of concern. Others explored their bots opportunistically (23\%) or did not substantively test them (16\%).

More rigorous testing did not appear to prevent deployment. Among bots that showed iterative refinement, limited evaluation, or opportunistic exploration, 53\%, 46\%, and 37\% were later used by non-authors. Overall, most educators tried their bots before release, but systematic and iterative testing was uncommon.

\subsubsection{More testing did not lead to better performance after deployment.}

End users held 902 conversations with 36 deployed bots. 
Overall, 75\% of bot responses were consistent with the behavior described in their system prompts. However, more testing was not associated with fewer failures after deployment. Bots whose builders ran more test-like conversations did not have lower failure rates (Spearman's $\rho = 0.05$, $p = .791$, $n = 36$). Failure rates also did not differ across the four testing modes (Kruskal-Wallis $p = .564$). This suggests that the amount of testing alone did not explain how well bots performed. We therefore examined whether educators had tested the kinds of situations that end users later brought to their bots.

\subsubsection{Testing covered routine use well, but missed most adversarial use.}

We coded builder and end-user messages using the same codebook (Section~3.4). Each user turn was coded as either \textit{desired} or \textit{adversarial} use, and adversarial turns were further coded by intent. For each bot, we then checked whether the situations seen during deployment had already appeared in the builder's own conversations. Figure~\ref{fig:lo3-perbot-coverage} compares the adversarial intents and bot failure types observed in builder and end-user conversations for each bot.

For \textit{desired} use, coverage was high. Across deployed bots, 91\% of desired-use bot--intent pairs seen in deployment had already appeared in builder conversations. In other words, educators had tested most of the routine ways their bots were later used. The pattern was very different for \textit{adversarial} use. End users probed 14 bots adversarially, but builders had done so for only 7 of them. Across these bots, end users produced 23 adversarial bot--intent pairs, including answer seeking, ghostwriting, boundary probing, jailbreak attempts, and detection evasion. Builders had previously tested only 6 of these 23 pairs (26\%), and four of those six were boundary probes. Thus, 74\% of the adversarial use patterns seen after deployment had not been tested beforehand.

These untested cases also included important failures. Answer dumping accounted for 38\% of end-user-side failures, compared with 4\% of builder-side failures. Among the answer-dumping failures, 92\% occurred on bots whose prompts already included an instruction not to give direct answers. Builders had tested answer seeking on only 2 of the 8 bots where end users later attempted it. Deployment also revealed platform failures, including hidden messages and timeouts, that had not appeared during builder testing.

\definecolor{successred}{RGB}{82, 120, 91}
\definecolor{lessongreen}{RGB}{164, 82, 72}

\section{Success and Lessons Learned}
 
\subsection{\textcolor{successred}{Success 1:} Learning-by-doing assignments extended into real-world use}

The assignment required educators to build a chatbot, but deploying it was optional. Even so, 36 of the 214 bots (17\%) reached real audiences, voluntarily and without compensation, and held 902 end-user conversations in which 75\% of bot responses behaved as their system prompts specified (LO3).

Continued use beyond the chatbot-building course provides a behavioral outcome measure for educator AI literacy programs. It may also benefit students. For assessment, most current professional development programs rely on attitude surveys and pre-post quizzes~\citep{lintner2024scales,chiu2024developing}, and self-reported competence correlates only weakly with demonstrated competence~\citep{zhang2026misalignment}. A deployed artifact triangulates both with a third source, what educators built and how it behaved with real users in their own teaching. This gives the measure the external validity that authentic assessment has long called for~\citep{wiggins1989teaching}. For students, instructor-configured tutors have produced larger learning gains~\citep{kestin2025ai,bastani2025generative}, so a guardrailed bot may serve students better than the commercial chatbot they would otherwise use.

Future programs can borrow this curriculum design by having the assignment produce an artifact that participating educators can use in their own teaching. This carries two advantages. First, deploying bots to real users in authentic educational settings gives the program a behavioral outcome measure at no additional cost. It also gives students an assistant shaped by their own instructor rather than a generic substitute.

\subsection{\textcolor{successred}{Success 2:} Shareable, interactive artifacts encouraged peer engagement}

Of the 214 bots, 63 (29\%) were used by fellow course participants, who held 125 peer-bot conversations (LO3). That is nearly twice the 36 bots that reached end users: colleagues tried each other's work more often than end users did. The bots that circulated addressed problems colleagues recognizably shared, such as syllabus and course-policy assistants, functioning as boundary objects for comparing practice. Educator-facing bots made up a third of the corpus (Figure~\ref{fig:lo1-overview}b).

Professional learning research consistently finds community formation difficult to engineer in online professional development~\citep{macia2016informal,lantzandersson2018twenty}. Here peer use emerged as a side effect of an artifact assignment, without design intent, suggesting shareable artifacts can do work that discussion forums are usually asked, and usually fail, to do.

Future programs need only two conditions to replicate this: artifacts that are one-click shareable, and assignments aimed at problems common across participants. Surfacing peer adoption to authors is a worthwhile third, plausibly a stronger motivator than a completion certificate.
 
\subsection{\textcolor{lessongreen}{Lesson 1:} Iterative testing did not emerge on its own and must be taught}

Most educators tested, but iterative refinement was uncommon. Iterative refinement, in which a builder tests, revises the prompt, and tests again, was the rarest mode: 14\% of bots and 23\% of educators reached it (Table~3). The median educator held three builder-side conversations, and 61\% of tested bots had all of theirs within one day and before deployment. These patterns suggest that iterative refinement needs more explicit instructional support (LO3).

Prior lab studies report the same pattern in AI chatbot authoring: non-experts judge a prompt from a single trial and overgeneralize from how they would instruct a person~\citep{zamfirescu2023johnny}; we observe it in trained educators with deployed artifacts. Course instruction on evaluating AI outputs (LO3) did not consistently translate into iterative prompt refinement, suggesting that evaluating a response and revising the prompt need distinct instructional support. Whether iteration reduces failures we cannot yet say, since testing mode was not associated with failure rate in our small deployed sample (Section~4.3). Lesson~2 suggests the reason lies in what educators iterated on, not whether they iterated.

The fix is partly curricular and partly infrastructural. On the curriculum side, teach iterative refinement as a named workflow (observe, revise, retest) and assess it from the trail it leaves: one before-and-after pair per bot, with the same test input, the failing response, the prompt edit, and the response after. On the platform side, Playlab keeps no accessible version history of prompt edits, so this trail exists only if educators record it by hand. The platform could store each prompt version with the responses it produced and suggest test cases from past failure logs, showing educators what changed and what still needs checking.

\subsection{\textcolor{lessongreen}{Lesson 2:} Awareness is not enough---teaching data-driven red-teaming for educational chatbot authoring}

Red-teaming means testing a system as an adversary would, probing for the inputs its designer did not intend \cite{feffer2024red}. Educators tested the right amount for the wrong distribution: their probes covered 91\% of the desired-use patterns end users later brought and 26\% of the adversarial ones (LO3). The gap mattered: of the answer-dumping failures end users triggered, 92\% occurred on bots whose prompts already forbade direct answers. Educators had written the guardrail but had not tried to break it.

Testing guidance for end-user programmers and educators is typically dispositional, an exhortation to test broadly~\citep{ribeiro2020beyond,shankar2024validators}, and a domain expert testing their own bot naturally simulates the cooperative student they teach for, not the one seeking a shortcut. Our data argue for reversing the order: derive adversarial categories from past cohorts' deployment data, then teach probing in exactly those. What transfers to other programs is not only our specific category list but the method: let the failure distribution write the red-teaming syllabus. Verification matters because prompt-level guardrails determine whether students gain or lose from AI tutors~\citep{bastani2025generative}.

This record spans 17,000+ messages from builders, peers, and end users conversing with educational AI chatbots. Its scale and richness give future research and teaching a context-relevant, evidence-based direction for red-teaming. Teacher training for AI tool authoring can now guide educators explicitly to test their bots against the adversarial uses end users actually attempted, answer seeking, ghostwriting, boundary probing, jailbreak attempts, and detection evasion, with one test conversation per adversarial category.
 
\section{Limitations and Future Work}

This work has several limitations. First, the course had no pre-test, so the record shows what educators did after training, not what training changed; the next cohort will add a baseline authoring task. Nor was there a post-test or self-report survey to triangulate the log-based findings against. The proposed changes are testable with the same instrumentation. Some gaps may belong to tooling rather than curriculum. Playlab keeps no version history of system-prompt edits, so iteration is hard to measure, and its anonymous links make it impossible to tell whether two conversations came from the same user.

\bibliography{aaai2027}

\end{document}